
\normalbaselineskip=12pt
\baselineskip=12pt
\magnification=1200
\hsize 15.0truecm \hoffset 1.0 truecm
\vsize 24.0truecm
\nopagenumbers
\headline={\ifnum \pageno=1 \hfil \else\hss\tenrm\folio\hss\fi}
\pageno=1
\font\fthreeb=cmbx10 scaled\magstep1
\def\lsim{\mathrel{\rlap{\lower4pt\hbox{\hskip1pt$\sim$}}
    \raise1pt\hbox{$<$}}}         
\def\gsim{\mathrel{\rlap{\lower4pt\hbox{\hskip1pt$\sim$}}
    \raise1pt\hbox{$>$}}}         


\centerline{\fthreeb  Higher twist corrections to Bjorken sum rule}
\vskip 36pt\centerline{ Mauro Anselmino}
\vskip 12pt
\centerline{\it Dipartimento di Fisica Teorica, Universit\`a di
Torino}
\centerline{\it and Istituto Nazionale di Fisica Nucleare,
Sezione di Torino}
\centerline{\it Via P. Giuria 1, I--10125 Torino, Italy}
\vskip 20pt
\centerline{ Francisco Caruso\footnote*{Also at Physics Institute of the
Universidade do Estado do Rio de Janeiro, 20550-013, Rio de Janeiro, Brazil}
and Eugene Levin\footnote{**}{On leave of absence from Theory Department
of Petersburg Nuclear Physics Institute, 188350, St. Petersburg, Gatchina,
Russia}}
\vskip 12pt
\centerline{\it Centro Brasileiro de Pesquisas F\'\i sicas/CNPq}
\centerline{\it Rua Dr. Xavier Sigaud 150, 22290-180, Rio de
Janeiro, Brazil}
\vskip 1.0in
\centerline{\bf ABSTRACT}
\vskip 12pt
Some higher twist corrections to the Bjorken sum rule are estimated
in the framework of a quark-diquark model of the nucleon. The parameters
of the model have been previously fixed by fitting the measured higher
twist corrections to the unpolarized structure function $F_2(x, Q^2)$.
The resulting corrections to the Bjorken sum rule turn out to
be negligible.
\vfill
\eject
\baselineskip 18pt plus 2pt minus 2pt
Several recent measurements of the polarized structure functions
$g_1^p(x)$ and $g_1^n(x)$ [1] both for protons and neutrons
have allowed the first tests of the fundamental Bjorken sum rule [2]:
$$
\int_0^1 dx~[g_1^p(x) - g_1^n(x)] = S_{Bj} = {a_3 \over 6} E_{NS}(Q^2)
\eqno(1)
$$
where $a_3 = 1.2573 \pm 0.0028$ is the neutron $\beta$-decay axial coupling.
The above equation holds on ground of isospin invariance, at leading twist
in the Operator Product Expansion; the coefficient function $E_{NS}$ has been
computed up to third order [3] in the strong coupling constant $\alpha_s$:
$$
E_{NS}(Q^2) =  1 - {\alpha_s \over \pi} - 3.58 \left( {\alpha_s \over \pi}
\right)^2 - 20.22 \left( {\alpha_s \over \pi} \right)^3
\eqno(2)
$$
where a number of three active flavours is assumed.

Data are available at $Q^2 = 2$ and 4.6 (GeV/c)$^2$ [1,4,5] for the L.H.S.
of Eq. (1): $0.152 \pm 0.014 \pm 0.021$ and $0.201 \pm 0.039 \pm 0.050$,
respectively.  At such values of $Q^2$ a comparison of Bjorken sum
rule with experiment should also take into account possible higher twist or
target mass corrections, not included in Eq. (1): $$
\int_0^1 dx~[g_1^p(x) - g_1^n(x)] = S_{Bj}(Q^2) + \delta S^{Bj}_{HT}
+ \delta S^{Bj}_{T} \,. \eqno(3)
$$

Some evaluations of target mass [6-8] and higher twist corrections can be
found in the literature [1], either in the framework of the
Drell-Hearn-Gerasimov sum rule [9] or of the QCD sum rules [6,10]. Some of
these corrections may be sizeable at $Q^2 \simeq$ 2 (GeV/c)$^2$, but the
final comparison with data shows no significant violation of the Bjorken
sum rule. Let us notice that reliable estimates of the higher
twist contributions are of fundamental importance in view of the recent
suggestions [7,11] to exploit the Bjorken sum rule (1) in order to extract
the value of the strong coupling constant $\alpha_s$.

To this purpose we consider here another possible evaluation of higher twist
corrections to $S_{Bj}$, in the framework of a quark-diquark model of the
nucleon. Diquarks originate from QCD colour forces and model correlations
between two quarks inside the nucleon which, at moderate $Q^2$ values,
interact collectively and behave as bound states; diquarks effectively
describe the attractive QCD forces between two quarks inside a baryon.
Evidence in support of the diquark model is given by a number of
phenomenological successes in the description of several physical phenomena,
such as: the behaviour of the DIS structure functions at Bjorken $x \to 1$,
the inclusive large $p_T$ deuteron production observed in 70 GeV/c $pp$
interaction at Serpukhov, the large $p_T$ baryon production in $pp$
interactions, the observed (but scarce) production of $\Delta^{++}$ at
large transverse momentum, and many other processes reviewed in Ref. [12].
It is also worthwhile mentioning that this approach has recently found some
further theoretical support in the Random Instanton Liquid Model for the
structure of QCD vacuum [13].

A most general version of the diquark model has been recently used to
describe, with excellent results, the higher twist contributions to the
unpolarized proton structure function $F_2(x,Q^2)$ [14]. Such fit has
led to values of the parameters of the model in good agreement with several
previous applications of simplified versions of the same model [12]. We would
like to stress that the diquark approach, being a physically motivated
self consistent model rather than an expansion in powers of $1/Q^2$,
has the unique feature of taking into account a full set of higher twist
corrections and not only the leading ones. Confident in its physical
content, we adopt here the same diquark model, with the same parameters, as
in Ref. [14]; in that our results are now genuine, parameter free predictions.

We recall that when probing the nucleon, in a quark-diquark state [14,15],
with the virtual photon in DIS one has three kinds of contributions:
the scattering off the single quarks outside the diquark,
the elastic scattering off the diquark, and the inelastic diquark
contribution, {\it i.e.}, the scattering off one of the quarks inside the
diquark. At large $Q^2$ values the diquark contributions, weighted by form
factors, vanish and one recovers the usual pure quark results.

The elastic diquark contributions have been studied in Ref. [15] and, for
the polarized structure function $g_1(x,Q^2)$ they read
$$
g_1^{(V)}(x,Q^2) = {1\over 4} e_V^2 \, \Delta V(x) \left[ \left(
2 + {Q^2 \over 2m_N^2 x^2} \right) (D_1 D_2 + {1 \over 2} Q^2 D_2 D_3)
- {Q^2 \over 4m_N^2 x^2} D_2^2 \right]
\eqno(4)
$$
where $D_{1,2,3}$ are form factors appearing in the most general coupling
of the photon to a spin 1 diquark; scalar diquarks obviously do not
contribute to the polarized structure functions. $\Delta V$ denotes the
difference between the number density of vector diquarks with spin
parallel and antiparallel to the nucleon spin.

Following the notations of Ref. [14] the higher twist diquark contributions
to $g_1$ for protons and neutrons are given by:
$$\eqalign{
\left[ g_1^{HT} \right]_p &= {1\over 4} \left( {16 \over 9} \Delta V_{uu}
+ {1 \over 9} \Delta V_{ud} \right) \left[ \left( 2 + {Q^2 \over 2m_N^2 x^2}
\right) (D_1 D_2 + {1 \over 2} Q^2 D_2 D_3) \right. \cr
&- \left. {Q^2 \over 4m_N^2 x^2} D_2^2 \right]
- {1 \over 2} \left( {4 \over 9} \Delta u_{_{V_{uu}}} +
{4 \over 9} \Delta u_{_{V_{ud}}} + {1 \over 9} \Delta d_{_{V_{ud}}}
\right) D_V^2 \cr}
\eqno(5)
$$
$$\eqalign{
\left[ g_1^{HT} \right]_n &= {1\over 4} \left( {1 \over 9} \Delta V_{uu}
+ {1 \over 9} \Delta V_{ud} \right) \left[ \left( 2 + {Q^2 \over 2m_N^2 x^2}
\right) (D_1 D_2 + {1 \over 2} Q^2 D_2 D_3) \right. \cr
&- \left. {Q^2 \over 4m_N^2 x^2}D_2^2 \right]
- {1 \over 2} \left( {1 \over 9} \Delta u_{_{V_{uu}}} +
{4 \over 9} \Delta u_{_{V_{ud}}} + {1 \over 9} \Delta d_{_{V_{ud}}}
\right) D_V^2 \cr}
\eqno(6)
$$
where the suffices to $V$ indicate the quark content of the vector diquark;
the negative contributions come from the scattering off the quarks inside the
diquarks, weighted by a factor $(1 - D_V^2)$, with $D_V$ related to the
diquark form factor. In Eq. (6) we have already used isospin relationships
and all distribution functions refer to the proton; $\Delta q$ is the usual
helicity density carried by quark $q$ and the suffices refer to the type of
diquark it comes from.

Subtracting Eqs. (6) from (5) yields
$$
\eqalign{
\left[ g_1^{HT} \right]_{p-n} &= {5\over 12} \Delta V_{uu}
\left[ \left( 2 + {Q^2 \over 2m_N^2 x^2}
\right) (D_1 D_2 + {1 \over 2} Q^2 D_2 D_3) - {Q^2 \over 4m_N^2 x^2}D_2^2
\right] \cr
&- {1 \over 6} \Delta u_{_{V_{uu}}} D_V^2 \,. \cr}
\eqno(7)
$$

In order to give numerical estimates we have to introduce explicit expressions
for the polarized distribution functions and the diquark form factors; the
valence quark content of the proton is assumed to be given by the flavour
and spin wave function
$$
\eqalign{
|p, S_z = \pm 1/2 \rangle &= \pm {1 \over 3} \left\{ \sin\Omega \, [
\sqrt 2 V^{\pm 1}_{(ud)} u^\mp - 2 V^{\pm 1}_{(uu)} d^\mp \right. \cr
& \left. + \sqrt 2 V^{0}_{(uu)} d^\pm - V^{0}_{(ud)} u^\pm ]
\mp 3 \cos\Omega \, S_{(ud)} u^\pm \right\} \cr}
\eqno(8)
$$
where $\sin^2\Omega$ and $\cos^2\Omega$ are, respectively, the probabilities
of finding a vector ($V$) or a scalar ($S$) diquark in the proton. According
to the above equation we have
$$
\eqalign{
\Delta V_{uu}(x) &= {4 \over 9} \sin^2\Omega\  f_{V_{uu}}(x) \cr
\Delta u_{_{V_{uu}}}(x) &= {8 \over 9} \sin^2\Omega\  f_{u_{_{V_{uu}}}}(x) \cr}
\eqno(9)
$$
where the $f$ distributions are normalized to $\int_0^1 dx~f(x) = 1$.

  From Eqs. (7), (9) and (3), upon integration over $x$ and taking, as in
Ref. [14], $D_1 = D_2 = D_V$, we obtain
$$
\eqalign{
\left[ \delta S^{Bj}_{HT} \right]_{diquarks} &=
{\sin^2\Omega \over 27} \Bigl\{ 6D_1^2 + 5 Q^2 D_1D_3 \cr
&+ {5Q^2 \over 4m_N^2} (D_1^2 + Q^2 D_1D_3) \int_0^1 {dx\over{x^2}}~
f_{V_{uu}}(x) \Bigr\} \,. \cr}
\eqno(10)
$$

Eq. (10) gives the quark-diquark model higher twist contributions to Bjorken
sum rule. We take the explicit expressions of the distribution function
$f_{V_{uu}}$ for a vector diquark and of the form factors from our previous
application of the same model to a fit of higher twist contributions to the
unpolarized structure function $F_2(x, Q^2)$ [14]:
$$\eqalign{
\sin^2\Omega &= 0.19 \cr
f_{V_{uu}} &= N x^{7.93}(1-x)^{3.32} \cr
D_1 = D_2 = D_V &= \left( {1.21 \over 1.21 + Q^2} \right)^2 \cr
D_3 &= {Q^2 \over m_N^4} D_1^2 \cr}
\eqno(11)
$$
where $N = 1/B(7.93; \, 3.32)$  is the normalization
constant ($B$ is the Euler beta function).

 From Eqs. (10) and (11) we obtain the numerical estimates at the
$Q^2$ values [in (GeV/c)$^2$] for which data are available:
$$
\eqalign{
\left[ \delta S^{Bj}_{HT}\,(Q^2=2.0) \right]_{diquarks} &\simeq
0.3~\times 10^{-2} \cr
\left[ \delta S^{Bj}_{HT}\,(Q^2=4.6) \right]_{diquarks} &\simeq
0.6~\times 10^{-3} \cr}
\eqno(12)
$$
The prediction of the model in the full range $1 \le Q^2 \le 10$ (GeV/c)$^2$
is shown in Fig 1.

It is interesting to compare the results given in Eqs. (12) not only with the
experimental values of $S_{Bj}$ [reported after Eq. (2)], but
also with the perturbative QCD higher order corrections of Eqs. (1) and (2),
namely with $a_3[E_{NS}(Q^2) -1]/6$. This definitely shows that the higher
twist contributions to the Bjorken sum rule, evaluated in the framework of the
quark-diquark model of the nucleon, turn out to be quite small.
Indeed our model yields a contribution which has an opposite sign with
respect to the pQCD corrections, but much smaller in magnitude:
at $Q^2 = 2$ and 4.6 (GeV/c)$^2$ we have, respectively,

\vskip 6pt
$$ {\left[ \delta S^{Bj}_{HT}\,(Q^2) \right]_{diquarks} \over {a_3
[1 - E_{NS}(Q^2)]/6}} \simeq 7\% \quad {\rm and} \quad 2\%. $$
\vskip 6pt

To better understand why it happens so, it might be instructive to compute
the diquark contribution to the first moment of $g_1^p(x)$ alone, {\it
i.e.}, $\Gamma_1^p = \int_0^1 dx~g_1^p(x)$. A sizeable diquark contribution
to $\Gamma_1^p$ would indicate that the tiny results obtained above for the
Bjorken sum, Eqs. (12), are only due to a strong cancellation between the
proton and the neutron contributions. However, an explicit calculation
shows that diquark\ contributions\ to\ $\Gamma_1^p$\ are\ comparable\ to
those found for the Bjorken sum rule, rejecting this possibility. We can
then safely conclude that the smallness of the diquark contributions to the
polarized structure functions is due to some intrinsic features of the
quark-diquark model of the nucleon: the strong $SU(6)$ violation
($\sin^2\Omega = 0.19$) favouring scalar diquarks; the mass scale of the
vector diquark form factor which turns out to be small, $Q_V^2 = 1.21$
(GeV/c)$^2$, corresponding to a large size, and the vector diquark $x$
distribution which is found to be peaked at $x \simeq 0.7$, suggesting that
vector diquarks consist of two almost uncorrelated quarks. In conclusion we
would like to repeat the main statement of our abstract, namely that
diquark contributions to the Bjorken sum rules turns out to be quite small.
We hope that our estimates will help in making the use of the Bjorken sum
rule in the experimental measurement of the strong coupling constant more
reliable.

\vskip 12pt
\noindent
{\bf Acknowledgements} - We thank M. Karliner and J. Feltesse
for many useful talks on the subject. One of us (M.A.) would like to thank
the members of the LAFEX at CBPF, where most of this work was done, for the
kind hospitality. F.C. and E.L. are grateful to CNPq of Brazil for a
financial support.

\vfill\eject
\noindent
{\bf Figure caption} - The value of $\left[ \delta S^{Bj}_{HT} \right]
_{diquarks}$, according to Eqs. (10) and (11) of the text, as a function
of $Q^2$.

\vskip 24pt
\noindent
{\bf References}
\vskip 12pt
\item{[ 1]} For a complete list of references and a comprehensive review
paper on the subject see, {\it e.g.}, M.~Anselmino, A.~Efremov and E.~Leader,
CERN preprint CERN-TH/7216/94, to appear in {\it Phys. Rep.}
\item{[ 2]} J.D. Bjorken, {\it Phys. Rev.} {\bf 148} (1966) 1467
\item{[ 3]} S.A. Larin and J.A.M. Vermaseren, {\it Phys. Lett.} {\bf B259}
(1991) 345; S.A. Larin, {\it Phys. Lett.} {\bf B334} (1994) 192
\item{[ 4]} B. Adeva {\it et al.}, {\it Phys. Lett.} {\bf B302} (1993) 533;
{\bf B320} (1994) 400
\item{[ 5]} D.L. Anthony {\it et al.}, {\it Phys. Rev. Lett.} {\bf 71}
(1993) 759
\item{[ 6]} I.I. Balitskii, V.M. Braun and A.V. Kolesnichenko, {\it Phys.
Lett.} {\bf B242} (1990) 245; {\bf B318} (1993) 648
\item{[ 7]} J. Ellis and M. Karliner, {\it Phys. Lett.} {\bf B313} (1993) 131
\item{[ 8]} H. Kawamura and T. Uematsu, preprint KUCP-76, hep-ph 9501368
\item{[ 9]} V.D. Burkert and B.L. Ioffe, CEBAF preprint,
CEBAF--PR--93--034, October 1993
\item{[10]} E. Stein, P. Gornicki, L. Mankiewicz, and A. Schaefer,
preprint UFTP 380/95, hep-ph 9502323
\item{[11]} J. Feltesse, {\it Proceedings of the XXVII
International Conference on High Energy Physics}, v. I, p. 65, Glasgow,
July 20-27, 1994; S.J. Brodsky, SLAC preprint, SLAC-PUB-95-6781
\item{[12]} M. Anselmino, S. Ekelin, S. Fredriksson, D.B. Lichtenberg and
E. Predazzi, {\it Rev. Mod. Phys.} {\bf 65} (1993) 1199
\item{[13]} T. Schafer, E.V. Shuryak and J.J.M. Verbaarschot, {\it Nucl.
Phys.}  {\bf B412} (1994) 143
\item{[14]} M. Anselmino, F. Caruso, J.R.T. de Mello, A. Penna Firme and
J. Soares, CBPF preprint CBPF-NF-024-94, 1994
\item{[15]} M. Anselmino, F. Caruso, E. Leader and J. Soares, {\it Z. Phys.}
{\bf C48} (1990) 689
\bye

{}From ANSELMINO@to.infn.it Fri May 12 04:00:18 1995
Return-Path: <ANSELMINO@to.infn.it>
Received: from to4ax8.to.infn.it by lafexSu1.lafex.cbpf.br (4.1/SMI-4.1)
	id AA12743; Fri, 12 May 95 03:59:32 EST
Date: Fri, 12 May 1995 8:59:50 +0300 (MET-DST)
{}From: ANSELMINO@to.infn.it
To: caruso@lafexSu1.lafex.cbpf.br
Cc: ANSELMINO@to.infn.it
Message-Id: <950512085950.3440155f@to.infn.it>
Subject: RE: fig. correta
Status: R

\normalbaselineskip=12pt
\baselineskip=12pt
\magnification=1200
\hsize 15.0truecm \hoffset 1.0 truecm
\vsize 24.0truecm
\nopagenumbers
\headline={\ifnum \pageno=1 \hfil \else\hss\tenrm\folio\hss\fi}
\pageno=1
\font\fthreeb=cmbx10 scaled\magstep1
\def\lsim{\mathrel{\rlap{\lower4pt\hbox{\hskip1pt$\sim$}}
    \raise1pt\hbox{$<$}}}         
\def\gsim{\mathrel{\rlap{\lower4pt\hbox{\hskip1pt$\sim$}}
    \raise1pt\hbox{$>$}}}         


\centerline{\fthreeb  Higher twist corrections to Bjorken sum rule}
\vskip 36pt\centerline{ Mauro Anselmino}
\vskip 12pt
\centerline{\it Dipartimento di Fisica Teorica, Universit\`a di
Torino}
\centerline{\it and Istituto Nazionale di Fisica Nucleare,
Sezione di Torino}
\centerline{\it Via P. Giuria 1, I--10125 Torino, Italy}
\vskip 20pt
\centerline{ Francisco Caruso\footnote*{Also at Physics Institute of the
Universidade do Estado do Rio de Janeiro, 20550-013, Rio de Janeiro, Brazil}
and Eugene Levin\footnote{**}{On leave of absence from Theory Department
of Petersburg Nuclear Physics Institute, 188350, St. Petersburg, Gatchina,
Russia}}
\vskip 12pt
\centerline{\it Centro Brasileiro de Pesquisas F\'\i sicas/CNPq}
\centerline{\it Rua Dr. Xavier Sigaud 150, 22290-180, Rio de
Janeiro, Brazil}
\vskip 1.0in
\centerline{\bf ABSTRACT}
\vskip 12pt
Some higher twist corrections to the Bjorken sum rule are estimated
in the framework of a quark-diquark model of the nucleon. The parameters
of the model have been previously fixed by fitting the measured higher
twist corrections to the unpolarized structure function $F_2(x, Q^2)$.
The resulting corrections to the Bjorken sum rule turn out to
be negligible.
\vfill
\eject
\baselineskip 18pt plus 2pt minus 2pt
Several recent measurements of the polarized structure functions
$g_1^p(x)$ and $g_1^n(x)$ [1] both for protons and neutrons
have allowed the first tests of the fundamental Bjorken sum rule [2]:
$$
\int_0^1 dx~[g_1^p(x) - g_1^n(x)] = S_{Bj} = {a_3 \over 6} E_{NS}(Q^2)
\eqno(1)
$$
where $a_3 = 1.2573 \pm 0.0028$ is the neutron $\beta$-decay axial coupling.
The above equation holds on ground of isospin invariance, at leading twist
in the Operator Product Expansion; the coefficient function $E_{NS}$ has been
computed up to third order [3] in the strong coupling constant $\alpha_s$:
$$
E_{NS}(Q^2) =  1 - {\alpha_s \over \pi} - 3.58 \left( {\alpha_s \over \pi}
\right)^2 - 20.22 \left( {\alpha_s \over \pi} \right)^3
\eqno(2)
$$
where a number of three active flavours is assumed.

Data are available at $Q^2 = 2$ and 4.6 (GeV/c)$^2$ [1,4,5] for the L.H.S.
of Eq. (1): $0.152 \pm 0.014 \pm 0.021$ and $0.201 \pm 0.039 \pm 0.050$,
respectively.  At such values of $Q^2$ a comparison of Bjorken sum
rule with experiment should also take into account possible higher twist or
target mass corrections, not included in Eq. (1): $$
\int_0^1 dx~[g_1^p(x) - g_1^n(x)] = S_{Bj}(Q^2) + \delta S^{Bj}_{HT}
+ \delta S^{Bj}_{T} \,. \eqno(3)
$$

Some evaluations of target mass [6-8] and higher twist corrections can be
found in the literature [1], either in the framework of the
Drell-Hearn-Gerasimov sum rule [9] or of the QCD sum rules [6,10]. Some of
these corrections may be sizeable at $Q^2 \simeq$ 2 (GeV/c)$^2$, but the
final comparison with data shows no significant violation of the Bjorken
sum rule. Let us notice that reliable estimates of the higher
twist contributions are of fundamental importance in view of the recent
suggestions [7,11] to exploit the Bjorken sum rule (1) in order to extract
the value of the strong coupling constant $\alpha_s$.

To this purpose we consider here another possible evaluation of higher twist
corrections to $S_{Bj}$, in the framework of a quark-diquark model of the
nucleon. Diquarks originate from QCD colour forces and model correlations
between two quarks inside the nucleon which, at moderate $Q^2$ values,
interact collectively and behave as bound states; diquarks effectively
describe the attractive QCD forces between two quarks inside a baryon.
Evidence in support of the diquark model is given by a number of
phenomenological successes in the description of several physical phenomena,
such as: the behaviour of the DIS structure functions at Bjorken $x \to 1$,
the inclusive large $p_T$ deuteron production observed in 70 GeV/c $pp$
interaction at Serpukhov, the large $p_T$ baryon production in $pp$
interactions, the observed (but scarce) production of $\Delta^{++}$ at
large transverse momentum, and many other processes reviewed in Ref. [12].
It is also worthwhile mentioning that this approach has recently found some
further theoretical support in the Random Instanton Liquid Model for the
structure of QCD vacuum [13].

A most general version of the diquark model has been recently used to
describe, with excellent results, the higher twist contributions to the
unpolarized proton structure function $F_2(x,Q^2)$ [14]. Such fit has
led to values of the parameters of the model in good agreement with several
previous applications of simplified versions of the same model [12]. We would
like to stress that the diquark approach, being a physically motivated
self consistent model rather than an expansion in powers of $1/Q^2$,
has the unique feature of taking into account a full set of higher twist
corrections and not only the leading ones. Confident in its physical
content, we adopt here the same diquark model, with the same parameters, as
in Ref. [14]; in that our results are now genuine, parameter free predictions.

We recall that when probing the nucleon, in a quark-diquark state [14,15],
with the virtual photon in DIS one has three kinds of contributions:
the scattering off the single quarks outside the diquark,
the elastic scattering off the diquark, and the inelastic diquark
contribution, {\it i.e.}, the scattering off one of the quarks inside the
diquark. At large $Q^2$ values the diquark contributions, weighted by form
factors, vanish and one recovers the usual pure quark results.

The elastic diquark contributions have been studied in Ref. [15] and, for
the polarized structure function $g_1(x,Q^2)$ they read
$$
g_1^{(V)}(x,Q^2) = {1\over 4} e_V^2 \, \Delta V(x) \left[ \left(
2 + {Q^2 \over 2m_N^2 x^2} \right) (D_1 D_2 + {1 \over 2} Q^2 D_2 D_3)
- {Q^2 \over 4m_N^2 x^2} D_2^2 \right]
\eqno(4)
$$
where $D_{1,2,3}$ are form factors appearing in the most general coupling
of the photon to a spin 1 diquark; scalar diquarks obviously do not
contribute to the polarized structure functions. $\Delta V$ denotes the
difference between the number density of vector diquarks with spin
parallel and antiparallel to the nucleon spin.

Following the notations of Ref. [14] the higher twist diquark contributions
to $g_1$ for protons and neutrons are given by:
$$\eqalign{
\left[ g_1^{HT} \right]_p &= {1\over 4} \left( {16 \over 9} \Delta V_{uu}
+ {1 \over 9} \Delta V_{ud} \right) \left[ \left( 2 + {Q^2 \over 2m_N^2 x^2}
\right) (D_1 D_2 + {1 \over 2} Q^2 D_2 D_3) \right. \cr
&- \left. {Q^2 \over 4m_N^2 x^2} D_2^2 \right]
- {1 \over 2} \left( {4 \over 9} \Delta u_{_{V_{uu}}} +
{4 \over 9} \Delta u_{_{V_{ud}}} + {1 \over 9} \Delta d_{_{V_{ud}}}
\right) D_V^2 \cr}
\eqno(5)
$$
$$\eqalign{
\left[ g_1^{HT} \right]_n &= {1\over 4} \left( {1 \over 9} \Delta V_{uu}
+ {1 \over 9} \Delta V_{ud} \right) \left[ \left( 2 + {Q^2 \over 2m_N^2 x^2}
\right) (D_1 D_2 + {1 \over 2} Q^2 D_2 D_3) \right. \cr
&- \left. {Q^2 \over 4m_N^2 x^2}D_2^2 \right]
- {1 \over 2} \left( {1 \over 9} \Delta u_{_{V_{uu}}} +
{4 \over 9} \Delta u_{_{V_{ud}}} + {1 \over 9} \Delta d_{_{V_{ud}}}
\right) D_V^2 \cr}
\eqno(6)
$$
where the suffices to $V$ indicate the quark content of the vector diquark;
the negative contributions come from the scattering off the quarks inside the
diquarks, weighted by a factor $(1 - D_V^2)$, with $D_V$ related to the
diquark form factor. In Eq. (6) we have already used isospin relationships
and all distribution functions refer to the proton; $\Delta q$ is the usual
helicity density carried by quark $q$ and the suffices refer to the type of
diquark it comes from.

Subtracting Eqs. (6) from (5) yields
$$
\eqalign{
\left[ g_1^{HT} \right]_{p-n} &= {5\over 12} \Delta V_{uu}
\left[ \left( 2 + {Q^2 \over 2m_N^2 x^2}
\right) (D_1 D_2 + {1 \over 2} Q^2 D_2 D_3) - {Q^2 \over 4m_N^2 x^2}D_2^2
\right] \cr
&- {1 \over 6} \Delta u_{_{V_{uu}}} D_V^2 \,. \cr}
\eqno(7)
$$

In order to give numerical estimates we have to introduce explicit expressions
for the polarized distribution functions and the diquark form factors; the
valence quark content of the proton is assumed to be given by the flavour
and spin wave function
$$
\eqalign{
|p, S_z = \pm 1/2 \rangle &= \pm {1 \over 3} \left\{ \sin\Omega \, [
\sqrt 2 V^{\pm 1}_{(ud)} u^\mp - 2 V^{\pm 1}_{(uu)} d^\mp \right. \cr
& \left. + \sqrt 2 V^{0}_{(uu)} d^\pm - V^{0}_{(ud)} u^\pm ]
\mp 3 \cos\Omega \, S_{(ud)} u^\pm \right\} \cr}
\eqno(8)
$$
where $\sin^2\Omega$ and $\cos^2\Omega$ are, respectively, the probabilities
of finding a vector ($V$) or a scalar ($S$) diquark in the proton. According
to the above equation we have
$$
\eqalign{
\Delta V_{uu}(x) &= {4 \over 9} \sin^2\Omega\  f_{V_{uu}}(x) \cr
\Delta u_{_{V_{uu}}}(x) &= {8 \over 9} \sin^2\Omega\  f_{u_{_{V_{uu}}}}(x) \cr}
\eqno(9)
$$
where the $f$ distributions are normalized to $\int_0^1 dx~f(x) = 1$.

 From Eqs. (7), (9) and (3), upon integration over $x$ and taking, as in
Ref. [14], $D_1 = D_2 = D_V$, we obtain
$$
\eqalign{
\left[ \delta S^{Bj}_{HT} \right]_{diquarks} &=
{\sin^2\Omega \over 27} \Bigl\{ 6D_1^2 + 5 Q^2 D_1D_3 \cr
&+ {5Q^2 \over 4m_N^2} (D_1^2 + Q^2 D_1D_3) \int_0^1 {dx\over{x^2}}~
f_{V_{uu}}(x) \Bigr\} \,. \cr}
\eqno(10)
$$

Eq. (10) gives the quark-diquark model higher twist contributions to Bjorken
sum rule. We take the explicit expressions of the distribution function
$f_{V_{uu}}$ for a vector diquark and of the form factors from our previous
application of the same model to a fit of higher twist contributions to the
unpolarized structure function $F_2(x, Q^2)$ [14]:
$$\eqalign{
\sin^2\Omega &= 0.19 \cr
f_{V_{uu}} &= N x^{7.93}(1-x)^{3.32} \cr
D_1 = D_2 = D_V &= \left( {1.21 \over 1.21 + Q^2} \right)^2 \cr
D_3 &= {Q^2 \over m_N^4} D_1^2 \cr}
\eqno(11)
$$
where $N = 1/B(7.93; \, 3.32)$  is the normalization
constant ($B$ is the Euler beta function).

 From Eqs. (10) and (11) we obtain the numerical estimates at the
$Q^2$ values [in (GeV/c)$^2$] for which data are available:
$$
\eqalign{
\left[ \delta S^{Bj}_{HT}\,(Q^2=2.0) \right]_{diquarks} &\simeq
0.3~\times 10^{-2} \cr
\left[ \delta S^{Bj}_{HT}\,(Q^2=4.6) \right]_{diquarks} &\simeq
0.6~\times 10^{-3} \cr}
\eqno(12)
$$
The prediction of the model in the full range $1 \le Q^2 \le 10$ (GeV/c)$^2$
is shown in Fig 1.

It is interesting to compare the results given in Eqs. (12) not only with the
experimental values of $S_{Bj}$ [reported after Eq. (2)], but
also with the perturbative QCD higher order corrections of Eqs. (1) and (2),
namely with $a_3[E_{NS}(Q^2) -1]/6$. This definitely shows that the higher
twist contributions to the Bjorken sum rule, evaluated in the framework of the
quark-diquark model of the nucleon, turn out to be quite small.
Indeed our model yields a contribution which has an opposite sign with
respect to the pQCD corrections, but much smaller in magnitude:
at $Q^2 = 2$ and 4.6 (GeV/c)$^2$ we have, respectively,

\vskip 6pt
$$ {\left[ \delta S^{Bj}_{HT}\,(Q^2) \right]_{diquarks} \over {a_3
[1 - E_{NS}(Q^2)]/6}} \simeq 7\% \quad {\rm and} \quad 2\%. $$
\vskip 6pt

To better understand why it happens so, it might be instructive to compute
the diquark contribution to the first moment of $g_1^p(x)$ alone, {\it
i.e.}, $\Gamma_1^p = \int_0^1 dx~g_1^p(x)$. A sizeable diquark contribution
to $\Gamma_1^p$ would indicate that the tiny results obtained above for the
Bjorken sum, Eqs. (12), are only due to a strong cancellation between the
proton and the neutron contributions. However, an explicit calculation
shows that diquark\ contributions\ to\ $\Gamma_1^p$\ are\ comparable\ to
those found for the Bjorken sum rule, rejecting this possibility. We can
then safely conclude that the smallness of the diquark contributions to the
polarized structure functions is due to some intrinsic features of the
quark-diquark model of the nucleon: the strong $SU(6)$ violation
($\sin^2\Omega = 0.19$) favouring scalar diquarks; the mass scale of the
vector diquark form factor which turns out to be small, $Q_V^2 = 1.21$
(GeV/c)$^2$, corresponding to a large size, and the vector diquark $x$
distribution which is found to be peaked at $x \simeq 0.7$, suggesting that
vector diquarks consist of two almost uncorrelated quarks. In conclusion we
would like to repeat the main statement of our abstract, namely that
diquark contributions to the Bjorken sum rules turns out to be quite small.
We hope that our estimates will help in making the use of the Bjorken sum
rule in the experimental measurement of the strong coupling constant more
reliable.

\vskip 12pt
\noindent
{\bf Acknowledgements} - We thank M. Karliner and J. Feltesse
for many useful talks on the subject. One of us (M.A.) would like to thank
the members of the LAFEX at CBPF, where most of this work was done, for the
kind hospitality. F.C. and E.L. are grateful to CNPq of Brazil for a
financial support.

\vfill\eject
\noindent
{\bf Figure caption} - The value of $\left[ \delta S^{Bj}_{HT} \right]
_{diquarks}$, according to Eqs. (10) and (11) of the text, as a function
of $Q^2$.

\vskip 24pt
\noindent
{\bf References}
\vskip 12pt
\item{[ 1]} For a complete list of references and a comprehensive review
paper on the subject see, {\it e.g.}, M.~Anselmino, A.~Efremov and E.~Leader,
CERN preprint CERN-TH/7216/94, to appear in {\it Phys. Rep.}
\item{[ 2]} J.D. Bjorken, {\it Phys. Rev.} {\bf 148} (1966) 1467
\item{[ 3]} S.A. Larin and J.A.M. Vermaseren, {\it Phys. Lett.} {\bf B259}
(1991) 345; S.A. Larin, {\it Phys. Lett.} {\bf B334} (1994) 192
\item{[ 4]} B. Adeva {\it et al.}, {\it Phys. Lett.} {\bf B302} (1993) 533;
{\bf B320} (1994) 400
\item{[ 5]} D.L. Anthony {\it et al.}, {\it Phys. Rev. Lett.} {\bf 71}
(1993) 759
\item{[ 6]} I.I. Balitskii, V.M. Braun and A.V. Kolesnichenko, {\it Phys.
Lett.} {\bf B242} (1990) 245; {\bf B318} (1993) 648
\item{[ 7]} J. Ellis and M. Karliner, {\it Phys. Lett.} {\bf B313} (1993) 131
\item{[ 8]} H. Kawamura and T. Uematsu, preprint KUCP-76, hep-ph 9501368
\item{[ 9]} V.D. Burkert and B.L. Ioffe, CEBAF preprint,
CEBAF--PR--93--034, October 1993
\item{[10]} E. Stein, P. Gornicki, L. Mankiewicz, and A. Schaefer,
preprint UFTP 380/95, hep-ph 9502323
\item{[11]} J. Feltesse, {\it Proceedings of the XXVII
International Conference on High Energy Physics}, v. I, p. 65, Glasgow,
July 20-27, 1994; S.J. Brodsky, SLAC preprint, SLAC-PUB-95-6781
\item{[12]} M. Anselmino, S. Ekelin, S. Fredriksson, D.B. Lichtenberg and
E. Predazzi, {\it Rev. Mod. Phys.} {\bf 65} (1993) 1199
\item{[13]} T. Schafer, E.V. Shuryak and J.J.M. Verbaarschot, {\it Nucl.
Phys.}  {\bf B412} (1994) 143
\item{[14]} M. Anselmino, F. Caruso, J.R.T. de Mello, A. Penna Firme and
J. Soares, CBPF preprint CBPF-NF-024-94, 1994
\item{[15]} M. Anselmino, F. Caruso, E. Leader and J. Soares, {\it Z. Phys.}
{\bf C48} (1990) 689
\bye